\documentclass[twocolumn,a4paper]{article}
\usepackage{graphics}
\begin{document}
\setlength{\unitlength}{1.0mm}
\title{Switching and symmetry breaking behaviour of discrete breathers
  in Josephson ladders}
\author{ Richard T Giles and Feodor V Kusmartsev
 \\ Department of Physics, Loughborough University,
 Loughborough, LE11 3TU, U.K.}
\date{April 4, 2000}
\maketitle
\textit{
  We investigate the roto-breathers recently observed in experiments
  on Josephson ladders subjected to a uniform transverse bias current.
  We describe the switching mechanism in which the number of rotating
  junctions increases. In the region close to switching we find that
  frequency locking, period doubling, quasi-periodic behaviour and
  symmetry breaking all occur. This suggests that a chaotic dynamic
  occurs in the switching process. Close to
  switching the induced flux increases sharply and clearly plays an
  important role in the switching mechanism. We also find three
  critical frequencies which are independent of the dissipation
  constant $\alpha$, provided that $\alpha$ is not too large.}

\vspace{10mm}
The recent discovery\cite{binder00,trias00} of discrete breathers in
Josephson ladder arrays (Fig.~\ref{fig:ladders}) driven by a
transverse dc bias current has shown not only that these localised
excitations exist but that they exhibit remarkable
behaviour, which is the subject of this paper.

The Josephson ladder has been studied theoretically for many years. In
the absence of a driving current the interaction between vortices has
been found to be exponential and this leads to the vortex density
exhibiting a devil's staircase as the magnetic field is
increased\cite{kardar84,giles99}. Quantum fluctuations\cite{kardar86},
meta-stable states\cite{mazo95} and inductance effects\cite{mazo96}
have been studied as has the interaction of
vortices with a transverse dc bias current\cite{kim97a,barahona98}.

In the case of a Josephson ladder, a roto-breather is a stable group
of vertical junctions rotating together
($\theta_j^\prime-\theta_j\approx\omega t$, see
Fig.~\ref{fig:ladders}). Until recently, the interest has focussed on
the case of a sinusoidal bias current, where it has been shown that
roto-breather solutions exist and that a discrete breather may even
include a chaotic trajectory\cite{martinez99}.

Roto-breathers should also be stable\cite{mackay98} in the
experimentally easier case of a ladder array subjected to a uniform dc
transverse driving current.  The case of rotation occurring at a
single vertical junction has been studied
numerically\cite{mazo99,flach99}.  Two experimental
groups\cite{binder00,trias00} have now independently confirmed the
existence of such solutions and discovered further unpredicted
behaviour. In an annular ladder\cite{binder00}
(Fig.~\ref{fig:ladders}a) and in a linear ladder\cite{trias00}
(Fig.~\ref{fig:ladders}b) breathers were observed with various numbers
of ``vertical'' (i.e. radial) junctions rotating. However the most
interesting features of the observations were the switching between
breathers and their \textit{symmetry breaking} behaviour. Once a
breather has been initialised it is stable, but if the bias current is
slowly decreased then, at some critical value of the current, the
number of rotating junctions switches to a larger number and a new
breather is formed.  Furthermore, \textit{the new breather may have a
  different centre of symmetry from the old breather despite the
  symmetry of the ladder itself and the symmetry of the driving
  currents about a single vertical junction}; this is particularly
remarkable in the case of the annular ladder where translational
invariance should be exact. The production of breathers with an even
number of rotating junctions is, by itself, a demonstration of
symmetry breaking.  The symmetry breaking effect was not mentioned by
the experimenters, presumably because it could be due to experimental
imperfections. However we show that it should also arise in a perfect
experiment, the apparent mechanism being the occurrence of chaotic
dynamics in the switching region.

First we construct a new model, similar to but significantly different
from those proposed previously\cite{mazo99,flach99,kim97,martinez99}.
The current through a junction is determined by the RCSJ model
\begin{equation}
\frac{I}{I_{c}} = \frac{d^2\varphi}{dt^2}+\alpha\frac{d\varphi}{dt}+
\sin \varphi
\end{equation}
where $I_{c}$ is the critical current and
\begin{equation}
\varphi=\Delta\theta-\frac{2\pi}{\Phi_0}\int\mathbf{A}.\mathrm{d}\mathbf{l}
\end{equation}
where $\Delta\theta$ is the change in superconducting order
parameter~$\theta$ across the junction and $\mathbf{A}$ is the vector
potential.  The ``vertical'' (i.e.
radial) junctions may differ in area from the ``horizontal'' junctions
by an anisotropy parameter
$\eta=I_{ch}/I_{cv}=C_h/C_v=R_v/R_h$ where $I_{ch}$ ($I_{cv}$), $C_h$
($C_v$) and $R_h$ ($R_v$) are, respectively, the critical current,
capacitance and resistance of a horizontal (vertical) junction.  From
Fig.~\ref{fig:ladders}a we see that there are three unknowns per
plaquette: $\theta_j$, $\theta_j^\prime$ and $f_j$, where $f_j$ is the
total flux threading the $j$th plaquette. To solve for these unknowns
we construct three equations per plaquette as follows. The first two
equations are obtained from current conservation at the top (inner)
and bottom (outer) rails. The third equation is obtained by making the
approximation that the induced flux $f_j-f_a$ (where $f_a$ is the
applied flux) is produced solely by the currents flowing around the
immediate perimeter of the $j$th plaquette (Fig.~\ref{fig:ladders}c):
\begin{equation}
\frac{f_j-f_a}{\Phi_0} =\frac{\beta_L}{8\pi}
\left(I_j^v+I_j^{h}-I_{j+1}^v-I_j^{\prime h}\right)
\label{eq:induction}
\end{equation}
where $\beta_L$ is an inductance parameter.  This last equation makes
the plausible assumption that the plaquettes are more or less square
and that it is only the currents flowing around the plaquette that
produce the induced field. This differs from, and for square
plaquettes is more accurate than, the common
assumption\cite{flach99,mazo99} that the induced field is proportional
to the loop (or ``mesh'') current circulating the plaquette.

We impose the initial conditions, mimicing the experiments, that at $t=0$,
$\theta_j=\theta_j^\prime=0$, $f_j=0$ and $I_j=0$ for all $j$.
This means that $\theta_j+\theta_j^\prime=0$ for all time, i.e. there
is a symmetry between the inner and outer rails.
Using Landau gauge we then have the following pair of coupled
differential equations for each plaquette:
\begin{eqnarray}
    \label{eq:-}
    \lefteqn{
    \left\{\frac{d}{dt^2}+\alpha\frac{d}{dt}\right\}\left(-\theta_{j-1}^-
+4\theta_j^- -\theta_{j+1}^-\right)=2I_j+\sin\theta_{j-1}^- } \nonumber \\
 & & -4\sin\theta_j^- +\sin\theta_{j+1}^-
     + \frac{8\pi}{\beta_L}(f_{j-1}-f_j)
\end{eqnarray}
\begin{eqnarray}
\lefteqn{
\left\{\frac{d}{dt^2}+\alpha\frac{d}{dt}\right\}
\left(-2\pi\eta f_j + (1-\eta)(\theta_{j+1}^- -\theta_j^-)\right) = \sin\theta_j^- }
\nonumber \\
 & & - \sin\theta_{j+1}^-+
\frac{8\pi}{\beta_L}(f_{j}-f_a) +2\eta\sin\chi_j^-
\label{eq:f}
\end{eqnarray}
where $\theta_j^-=\theta_j^\prime-\theta_j$
and $\chi_j^-=\frac{1}{2}(\theta^-_{j+1}-\theta^-_j+2\pi f_j)$.

\begin{figure}[htbp]
\begin{picture}(80,80)
\put(10,3){\makebox(60,60){\includegraphics[80mm,113mm][132mm,164mm]
    {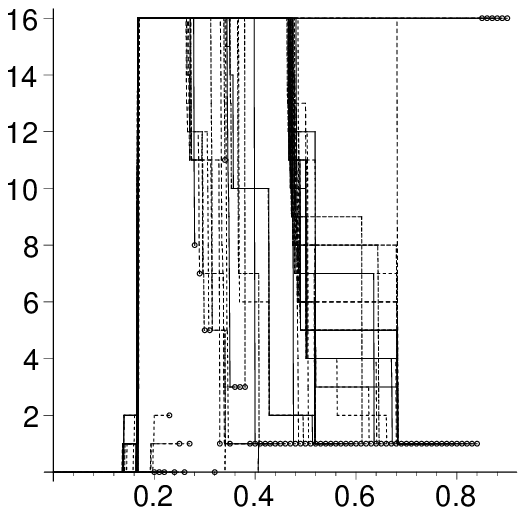}}}
\put(30,3){Bias current $I_B$}
\put(9,10){\rotatebox{90}{Number of rotating junctions, $N_R$}}
\end{picture}
\caption{\label{fig:ramp}Result of smoothly ramping down the bias
  current. Each small circle marks the start of a new ramp.}
\end{figure}

We now focus on determining whether or not our model exhibits the
interesting switching and symmetry breaking behaviour observed in the
data of Binder $et$ $al$\cite{binder00}. All parameter values are
chosen to mimic the experimental setup i.e. $\eta=0.44$,
$\beta_L=2.7$, $\alpha=0.07$ and $f_a=0$. Let
\begin{equation}
I_j=\left\{ \begin{array}{ll}
    I_B + I_\Delta & \mbox{ if $j=0$} \\
    I_B            & \mbox{ otherwise}
    \end{array}
\right.
\end{equation}
where $I_B$ is called the bias current.  Again following the
experiment (Fig.~\ref{fig:ladders}a), we slowly increase $I_\Delta$
while keeping $I_B=0$ until rotation starts at site~$j=0$. $I_\Delta$
is then slowly decreased while at the same time increasing $I_B$ to
keep $I_0$ constant. When $I_B$ has reached the desired value it is
then held fixed while $I_\Delta$ is slowly reduced to zero. Finally
$I_B$ is slowly reduced while keeping $I_\Delta=0$. Note that the
dynamical equations, initial conditions and injected currents are
exactly symmetrical about site $j=0$. One would expect only solutions
which are also symmetrical about~$j=0$.  However, like the
experiments, the simulations also produce breathers which are
\textit{not symmetrical} about~$j=0$.  Fig.~\ref{fig:ramp} shows how
the number~$N_R$ of rotating junctions changes as the bias
current~$I_B$ is slowly reduced from various starting values. In
agreement with experiment, breathers with \textit{even} $N_R$ are
commonly produced (these \textit{cannot} be symmetrical about~$j=0$),
and $N_R$ switches to larger and larger values until eventually all
junctions are rotating. While we have found that similar behaviour
occurs in a previously published model\cite{flach99,mazo99}, our model gives
considerably better agreement with the experimentally observed
switching currents, thus indicating the importance of inductance
effects in the switching mechanism. We believe the origin of the
symmetry breaking is the occurrence of chaotic dynamics in the
switching region.

The chaotic nature of the switching is further suggested by the fact
that when the same computer code which produced Fig.~\ref{fig:ramp} is
performed on a different computer (different floating point processor)
we see significant changes in the switching behaviour of nearly all
trajectories although the overall pattern remains identical. The
minute differences in the handling of floating point numbers are
amplified in the chaotic regime to produce significantly altered
trajectories.

\begin{figure}[htbp]
\begin{picture}(80,80)
\put(10,3){\makebox(60,60){\includegraphics[77mm,112mm][133mm,166mm]
    {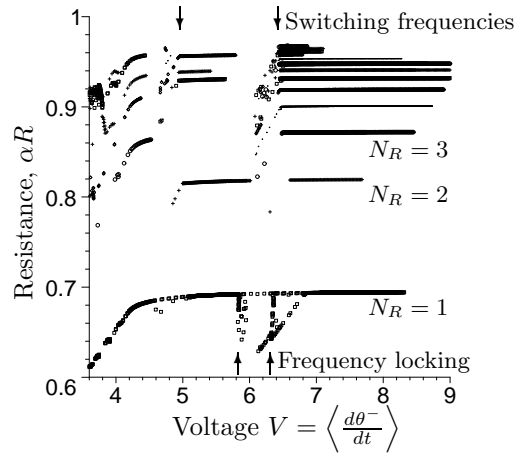}}}
\put(30,2){Voltage $V=\left\langle\frac{d\theta^-}{dt}\right\rangle$}
\put(9,20){\rotatebox{90}{Resistance, $\alpha R$}}
\put(56,18){\small $N_R=1$}
\put(56,33){\small $N_R=2$}
\put(56,39){\small $N_R=3$}
\put(43,10){\vector(0,1){3}}
\put(38.7,10){\vector(0,1){3}}
\put(44,11){\small Frequency locking}
\put(44,59){\vector(0,-1){3}}
\put(31,59){\vector(0,-1){3}}
\put(45,56){\small Switching frequencies}
\end{picture}
\caption{\label{fig:vi}Resistance~$\alpha R$ (where $R=V/I_B$) plotted against
  voltage~$V=\left\langle d\theta^-/dt\right\rangle $ for the data shown in
  Fig.~\ref{fig:ramp}. Note the critical switching frequencies and
  frequency locked states.}
\end{figure}

Fig.~\ref{fig:vi} shows the voltage-current characteristics obtained
from the same simulations used in producing Fig.~\ref{fig:ramp}.  At
high frequencies ($V=\left\langle d\theta^-/dt\right\rangle>4.3$) most
of the breathers show purely resistive behaviour (i.e.
$V=\left\langle d\theta^-/dt\right\rangle =I_B R$), the value of the
resistance~$R$ increasing with $N_R$ according to an expression
deduced from experiment\cite{binder00}:
\begin{equation}
\alpha R\approx 1/(1+\eta/N_R)
\label{eq:R}
\end{equation}
This is the relationship expected if the
current through each rotating vertical junction is $\alpha V$ and the four
horizontal junctions surrounding the rotating region each carry
current~$\frac{1}{2}\eta\alpha V$ (to satisfy Kirchoff's rule). For
$N_R=1$ this expression was derived in ref.~\cite{flach99}. We find
that no such resistive behaviour occurs for $V<4.3$.

\begin{figure}[htbp]
\begin{picture}(50,50)
\put(19,0){\makebox(0,0)[bl]{\includegraphics[84mm,116mm][137mm,157mm]
    {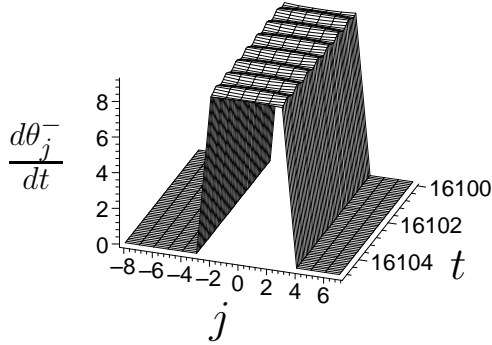}}}
\put(67,4){\LARGE$t$}
\put(35,-3){\LARGE$j$}
\put(8,18){\LARGE$\frac{d\theta_j^-}{dt}$}
\end{picture}
\caption{\label{fig:breather}An example of a resistive breather far
  from switching.  $d\theta^-_j/dt$ is plotted against $t$ for all
  $j$. Note that six junctions ($-2\le j\le 3$) are rotating and that
  the breather is \textit{not symmetrical} about the injection site
  $j=0$. The breather is periodic and exactly symmetrical about its
  centre. The peak flux (not shown) is small ($\sim 0.06$).}
\end{figure}

A typical example of a resistive breather (far away from any switching
region) is shown in
Fig.~\ref{fig:breather}.  Although the breather is not symmetric
about~$j=0$ it shows exact symmetry about the midpoint of the rotating
region and also appears to be exactly periodic, the period being two
revolutions of a vertical junction (we call this ``period 2''). The
non-rotating junctions oscillate, the amplitude decreasing
exponentially with distance from the rotating region.  Far away from
switching the magnitude
of the flux~$f_j$ is everywhere small ($<0.1$) and is limited to the
edges of the rotating region.

Fig.~\ref{fig:vi} shows that there are at least two critical
frequencies, $\left\langle d\theta^-/dt\right\rangle =6.5$ and
$\left\langle d\theta^-/dt\right\rangle=5.0$, at which breathers
become unstable and switching occurs. As the current is reduced the
frequency falls until it reaches the critical value, at which point
the breather becomes unstable and switches to a larger breather with a
larger resistance (Eq.~(\ref{eq:R})) and therefore a higher frequency.
The process then repeats as the current continues to be ramped down.
We identify the larger of these two critical frequencies with that
observed experimentally\cite{binder00}.  Although, at fixed~$I_B$, the
frequency of rotation depends on the dissipation constant~$\alpha$, we
find that these two critical frequencies are more or less independent
of $\alpha$, as is the critical frequency below which breathers cease
to show constant resistance. This of course breaks down when the bias
current required to achieve the upper critical frequency exceeds
unity, i.e. when $\alpha>0.15$.

Note also that in the above simulations: (i) no breathers are seen
for $I_B>0.85$, (ii) only the single site breather is seen for
$0.68<I_B<0.85$ and (iii) all rotating junctions normally stop
together when the current $I_B$ is reduced below 0.17, although
sometimes a new breather containing only a few rotating junctions is
produced which is then destroyed when $I_B$ falls below 0.14.

\begin{figure}[htbp]
\begin{picture}(50,50)
\put(19,0){\makebox(0,0)[bl]{\includegraphics[80mm,113mm][142mm,159mm]
    {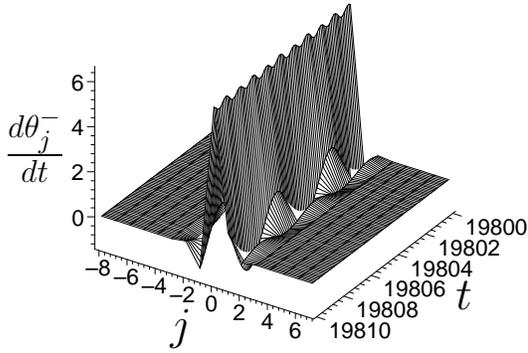}}}
\put(71,4){\LARGE$t$}
\put(33,0){\LARGE$j$}
\put(11,23){\LARGE$\frac{d\theta_j^-}{dt}$}
\end{picture}
\caption{\label{fig:quasi} The quasi-periodic breather (approximately
  period 8) which occurs near $I_B=0.6892$, just above the bias
  current at which switching to multi-site breathers occurs. It is
  frequency-locked at $\left\langle d\theta^-/dt \right\rangle =
  6.32$.  Note that $d\theta^-_j/dt$ is nearly anti-symmetric about
  $j=0$. The maximum flux (not shown) is large (0.5) and increases
  still further at switching.}
\end{figure}

As well as resistive period 2 breathers we have found
\textit{frequency-locked} breathers. In fact the usual $N_R=1$
resistive breather makes a transition to a frequency-locked breather
as the bias current is reduced below $I_B=0.699$. At this point the
period doubles, and the breather becomes \textit{asymmetric} and jumps
to a higher rotation frequency $\left\langle d\theta^-/dt\right\rangle
=6.32$. At $I_B=0.6895$ the period doubles again and at $I_B=0.6892$
it becomes quasi-periodic (Fig.~\ref{fig:quasi}).  Switching to
multi-site breathers occurs at $I_B\approx 0.6891$.  Such frequency
locked breathers have not been reported experimentally but have
probably been overlooked as they are only stable in narrow current
intervals.

While the maximum flux $f^{max}$ is normally small ($f^{max}<0.1$), it
sharply increase at switching to $f^{max}\sim 1$. It would appear
that an important part of the switching mechanism is the instability
caused by having a large flux concentrated in a small area. The
importance of the flux in the switching process is further confirmed
by the fact that when~$I_B$ has finally been reduced all the way to
zero we find that the ladder may contain one or more \textit{stable
  vortex-antivortex pairs}.

The occurrence of frequency locking, period doubling, quasi-periodic
behaviour and symmetry breaking as switching is approached strongly
suggests that the origin of the symmetry breaking is the occurrence of
chaotic dynamics in the switching region.  From the theory of chaotic
dynamics it is known that two or more coupled Josephson junctions (or,
equivalently, two or more coupled pendula) may exhibit three types of
motion: oscillatory, rotational and chaotic. Chaotic motion arises for
particular initial conditions. An analogous situation arises in
non-linear coupled lattices which also display both breathers and chaotic
dynamics\cite{flach98}. In a large array of coupled Josephson
junctions (or, analagously, an array of coupled pendula) we must also
expect these three types of motion. While roto-breathers are the most
characteristic stable solutions it appears that the switching between
breathers occurs only in the chaotic regime.  Of course, the chaotic
dynamics has an influence on the selection of breather type. Any small
difference in the initial conditions leads to a completely different
breather. Thus, in this chaotic regime, any small experimental
imperfections or small inaccuracies in a computer simulation may lead
to different chaotic trajectories and hence different roto-breathers.
Symmetry breaking arises if the perturbation itself breaks symmetry.

We conclude that our model exhibits most of the main features
\cite{footnote} of
the roto-breathers recently observed in Josephson ladder arrays, sheds
light on the switching mechanism and predicts further observable
behaviour. The occurrence of frequency locking, period doubling,
quasi-periodic behaviour and symmetry breaking in the switching region
suggests that switching occurs in the chaotic regime.  We find that
the maximum flux increases sharply in the switching region and that
the flux clearly plays an important role in the switching mechanism.
We also find two critical switching frequencies and a critical
frequency below which no resistive behaviour is observed. All three
frequencies are approximately independent of the dissipation
constant~$\alpha$ (for $\alpha<0.15$),

\vspace{5mm} We are grateful to A.V.~Ustinov and P.~Binder for kindly
giving us their data and explaining it prior to publication.
We are also grateful for discussions with M.V.~Fistul and
S.~Flach and for the hospitality of the Max-Planck-Institut, Dresden
and the Universit\"{a}t Erlangen-N\"{u}rnberg.


\begin{thebibliography}{10}
\bibitem{footnote}
Ref.~\cite{binder00} finds that breathers with even~$N_R$ are most common in
  the annular ladder. In agreement with this we find that, when switching,
  breathers with $N_R=2$ or $N_R=4$ have a strong tendency to grow
  symmetrically thus maintaining even $N_R$. No such effect is seen for
  breathers with $N_R=1$ or $N_R=3$. This is strongly suggestive of a parity
  effect, i.e. a difference in behaviour according to whether $N_R$ is even or
  odd.
\end{thebibliography}

\begin{figure}[htbp]
\begin{center}
\resizebox{60mm}{!}
{\includegraphics*[50mm,50mm][170mm,210mm]{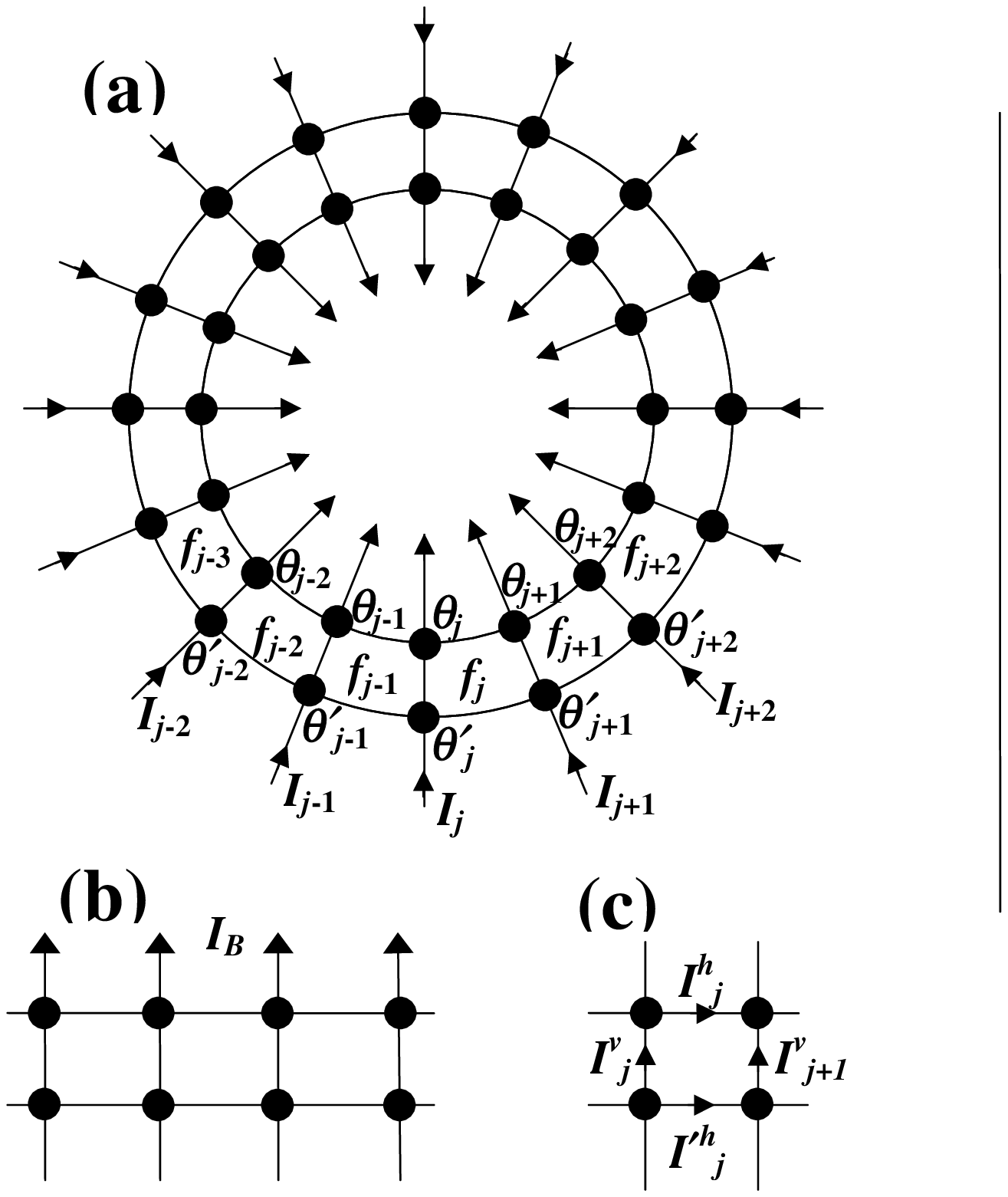}}
\end{center}
\caption{\label{fig:ladders}(a) An annular ladder subjected to a
  uniform transverse bias current~$I_B$. Each circle represents a
  superconducting island. Each link between islands represents a
  Josephson junction. $\theta$ is the phase of the superconducting
  order parameter and $f$ is the flux threading a plaquette. (b) A
  linear ladder. (c) Explanation of the notation used in
  Eq.~(\ref{eq:induction}).}
\label{fig:ladders}
\end{figure}

\end{document}